\documentclass[sn-aps]{sn-jnl}

\usepackage{graphicx}%
\usepackage{multirow}%
\usepackage{amsmath,amssymb,amsfonts}%
\usepackage{amsthm}%
\usepackage{mathrsfs}%
\usepackage{tensor}
\usepackage[title]{appendix}%
\usepackage{xcolor}%
\usepackage{textcomp}%
\usepackage{manyfoot}%
\usepackage{booktabs}%
\usepackage{algorithm}%
\usepackage{algorithmicx}%
\usepackage{algpseudocode}%
\usepackage{listings}
\usepackage{hyperref}
\hypersetup{
	colorlinks=true,       
	linkcolor=blue,        
	citecolor=blue,        
	filecolor=blue,     
	urlcolor=black         
}

\theoremstyle{thmstyleone}%
\theoremstyle{thmstyletwo}%

\theoremstyle{thmstylethree}%

\begin{document}

\title[Comoving curvature perturbation in Jordan and Einstein frames]{Comoving curvature perturbation in Jordan and Einstein frames}


\author[1]{\fnm{Jos\'e Jaime} \sur{Terente D\'iaz}}

\author[1]{\fnm{Mindaugas} \sur{Kar\v{c}iauskas}}


\affil[1]{\orgdiv{Departamento de F\'isica Te\'orica}, \orgname{Universidad Complutense de Madrid}, \orgaddress{\postcode{E-28040}, \state{Madrid}, \country{Spain}}}




\abstract{In the context of $F(\phi)R$ models of gravity, the conformal invariance of the curvature perturbation on uniform-field slices has been already demonstrated in several publications. In this work we study the curvature perturbation $\mathcal{R}$ defined on hypersurfaces that comove with the effective fluid whose energy-momentum tensor is covariantly conserved. We derive the expressions of $\mathcal{R}$ at first order in perturbations in the Jordan and Einstein frames and relate the two. Generically $\mathcal{R}$ is not conformally invariant, but it is on sufficiently large scales during slow-roll inflation. Using our results we also rederive the expressions for inflation observables in the Jordan frame.}

\keywords{Modified gravity, inflation, conformal frames, comoving slicing}



\maketitle

\section{Introduction}\label{sec1}
Over the last decades, mostly since the discovery of the accelerating expansion of the Universe \cite{SupernovaSearchTeam:1998fmf,SupernovaCosmologyProject:1998vns}, many theories of modified gravity have become an essential framework to explore the early and late history of the Universe, the primary example of the former being the paradigm of inflation \cite{Starobinsky:1979ty,Starobinsky:1980te,Guth:1980zm,Linde:1981mu}. Scalar-tensor theories, in particular, are among the most frequently used gravity models that have been applied to study inflation \cite{Clifton:2011jh,Nojiri:2017ncd,Futamase:1987ua,Fakir:1990iu,Bezrukov:2007ep,Cheong:2021vdb,Kodama:2021yrm}. These theories lead into the notion of conformal frames \cite{Dabrowski:2008kx} and raise the question of how to interpret them physically \cite{Faraoni:1999hp,Flanagan:2004bz,Deruelle:2010ht,Karam:2017zno,Rondeau:2017xck,Azri:2018gsz}.

In addition to those models of modified gravity, the inflation paradigm has strengthened the development of the theory of cosmological perturbations \cite{Kodama:1984ziu,Mukhanov:1990me,Malik:2008im}. Not in vain, one of its most striking predictions is the evolution of classical perturbations from quantum vacuum fluctuations into the large scale structures in the Universe \cite{Lyth:2009zz}. An important quantity in the inflationary cosmology is the curvature perturbation, often formulated on comoving or uniform-density time slicings. The perturbation defined on those slicings has been shown to become constant once the corresponding modes leave the horizon if the non-adiabatic pressure vanishes \cite{Weinberg:2003sw}. This was proven to hold non-perturbatively in Ref.~\cite{Lyth:2004gb}, even when modifications of Einstein's gravity are included.\footnote{Assuming the separate universes approach holds.} These modifications may generally be moved to the RHS of the metric field equations and interpreted as contributions to an effective energy-momentum tensor $\tilde{T}_{\mu\nu}$ \cite{Pimentel:1989bm,Faraoni:2018qdr}. The metric field equations take on the form of the Einstein ones in General Relativity (GR), and the effective energy-momentum tensor is covariantly conserved by virtue of Bianchi's identity \cite{Carroll:1997ar}.

In the context of scalar-tensor theories, cosmological perturbations have been extensively examined in the Jordan frame \cite{Hwang:1991aj,Makino:1991sg,Fakir:1992cg,Hwang:1996xh,Hwang:2001qk}, which is the frame where the gravitational sector of the action includes a non-minimal coupling to gravity and the matter sector couples to gravity only minimally. This analysis has been also performed in the post-inflationary universe, in the presence of matter fields \cite{Chiba:2013mha}. In those and other works \cite{Makino:1991sg,Fakir:1992cg,Chiba:2008ia,Gong:2011qe,Kubota:2011re} it is usually claimed that the curvature perturbation is conformally equivalent on comoving slicings \cite{Chiba:2008ia,Chiba:2013mha,Gong:2011qe}. This has been shown to be exactly true (on all length scales and at any expansion rate as we emphasise later, in Eq.~\eqref{eq:conf-trans-psi-deltaphi}) on uniform-field slicings for single scalar field models. Those are the slicings that comove with the scalar field fluid (see e.g. Eq.~(A.20) in Ref.~\cite{Chiba:2008ia}, and Eq.~(2.10) in Ref.~\cite{Gong:2011qe}, or discussion in the paragraph below Eq.~(2.3) in the same reference).\

One can also define the curvature perturbation on slices that comove with the effective fluid. We call this perturbation $\mathcal{R}$. It is to this perturbation that the results of Ref.~\cite{Lyth:2004gb} apply directly. With this definition of the \emph{comoving} slicing it does not generally coincide with the uniform-field slicing. This was clearly demonstrated in Ref.~\cite{Gong:2011qe} in the case of $f(R)$ gravity. The same conclusion holds for $F(\phi)R$ gravity, as we show it explicitly in this work. Indeed, the presence of gravity modifications in the effective energy-momentum tensor prevents the off-diagonal components $\tensor{\tilde{T}}{^{0}_{i}}$ (the momentum density) from vanishing when the perturbations of the non-minimally coupled scalar field $\phi$ (or $F\equiv f_{,R}$ in $f(R)$ gravity) are set to zero. Since all the perturbations defined on the uniform-field slicing are conformally invariant \cite{Gong:2011qe}, $\mathcal{R}$ will not be equivalent in different conformally related frames.\  

Given the relevance of the comoving curvature perturbation $\mathcal{R}$ in relating observations of the Cosmic Microwave Background (CMB) and the Large Scale Structure (LSS) with the properties of inflation \cite{Baumann:2009ds,ParticleDataGroup:2022pth}, we address the issue of its conformal invariance in $F(\phi)R$ theories of gravity at first order in perturbation theory, and derive the exact conformal relation between $\mathcal{R}$ in Jordan and Einstein frames (the latter being the frame in which the gravitational sector of the action corresponds to the Einstein-Hilbert action of GR). When applying the results of this analysis to slow-roll inflation, we use some of the results from our previous work \cite{Karciauskas:2022jzd}, where we derived the slow-roll approximation \cite{Liddle:1994dx} in the Jordan frame.

In the present work, we use geometrical units where $c=\hbar=M_{\mathrm{Pl}}=1$, $M_{\mathrm{Pl}} = (8\pi G)^{-1/2}$, and $G$ is the Newton's gravitational constant. We also adopt the `mostly positive' signature of the metric.

\section{Conformal frames and linear perturbations}\label{sec:2}
Conformal frames are naturally applied in scalar-tensor theories of gravity. In this section, we review the action of a field $\phi$ non-minimally coupled to gravity and introduce the conformal transformation of the metric that allows to write the action in the more familiar Einstein frame form. We begin by discussing homogeneous manifolds and move to linear perturbations about a spatially flat Friedman-Lema\^{i}tre-Robertson-Walker (FLRW) metric. To properly introduce the comoving curvature perturbation at first order in perturbation theory in the next sections, we re-examine how the scalar perturbations of the energy-momentum tensor (the scalar momentum in particular) and of the metric transform under gauge transformations. Finally, we summarise the formulae developed in Ref.~\cite{Karciauskas:2022jzd} for the $F(\phi)R$ theory of gravity needed to analyse the conformal relation between the linear comoving curvature perturbation in Jordan and Einstein frames during slow-roll, as well as the corresponding approximations.

\subsection{The conformal transformation}
To begin with, we consider a four dimensional spacetime manifold that is being foliated by a family of space-like hypersurfaces. Such a foliation or slicing is performed by adopting a coordinate time $t$ on the manifold \cite{Gourgoulhon:2007ue}. While $t$ is kept fixed, the proper time $\tau$ depends on the metric: \begin{equation} \textrm{d}\tau \equiv \mathcal{N} \textrm{d}t~,\end{equation} where $\mathcal{N}$ is the lapse function, which relates the coordinate time to the physical time $\tau$ measured by some fiducial observer \cite{Gourgoulhon:2007ue}.

We now assume FLRW spacetime manifolds, meaning that the slices are homogeneous and isotropic. The threading is chosen so that the shift vector vanishes, which simplifies the expressions. The resulting spatially flat FLRW metric is: \begin{equation}\label{eq:line-element-homogeneous-isotropic} \textrm{d}s^2 = -\mathcal{N}^2 \textrm{d}t^2+a^2(t) \delta_{ij} \textrm{d}x^{i} \textrm{d}x^{j}~,\end{equation} where $a(t)$ refers to the scale factor that parametrises the relative variation of the proper volume element $\delta \mathcal{V}\propto a^{3}$ of the space-like hypersurfaces over time. The volume expansion rate is given by the Hubble parameter \cite{Karciauskas:2022jzd}: \begin{equation}\label{eq:def-Hubble-parameter-proper-vol} H \equiv \frac{1}{3\delta \mathcal{V}}\frac{\textrm{d}\delta\mathcal{V}}{\textrm{d}\tau}~,\end{equation} which is equal to $-1/3$ of the trace of the extrinsic curvature tensor associated with the constant time hypersurface. In terms of the coordinate time: \begin{equation}\label{eq:Hubble-param-in-terms-of-a} H(t) = \frac{\dot a(t)}{\mathcal{N}a(t)}~,\end{equation} where overdots denote derivatives with respect to the coordinate time. 

We can use the same notation for the time derivative, i.e. the overdot, irrespective of the frame (metric) as we keep the coordinates of the homogeneous manifold fixed. However, the metric coordinate components (the lapse function and the scale factor) are altered by a transformation of the metric, and the proper time and proper volume shall differ in different frames. A particular transformation of the metric, which results in a rescaling of the lapse function and the scale factor, is the conformal one \cite{Dabrowski:2008kx}: \begin{equation} \label{eq:conf-metric}\hat{g}_{\mu\nu}(x) = \Omega^2(x) g_{\mu\nu}(x)~,\end{equation} where $\Omega^2(x)$ is the conformal factor. When applied to the homogeneous and isotropic metric \eqref{eq:line-element-homogeneous-isotropic}, the conformal transformation yields: \begin{align}\label{eq:ref-N-only-for-that}&\hat{\mathcal{N}} = \Omega \mathcal{N}~,\\
\label{eq:ref-a-only-for-that}&\hat{a} = \Omega a~,\end{align} where $\Omega = \Omega(t)$. Although the slicing of the spacetime or, equivalently, the choice of the time coordinate, is arbitrary, for convenience we fix the slicing of the unperturbed universe such that $\mathcal{N} = 1$ in the Jordan frame (without the loss of generality), following Ref.~\cite{Karciauskas:2022jzd}. The expressions are then simplified in that frame, which we shall introduce later. This implies that $\hat{\mathcal{N}} = \Omega$. 

When it comes to perturbations, we turn our attention to the scalar ones about the spatially flat FLRW background metric in particular. The metric is perturbed such that the full one is: \begin{equation}\label{eq:perturbed-metric-generic} g_{\mu\nu}(x) = \bar{g}_{\mu\nu}(t) + \delta g_{\mu\nu}(x)~,\end{equation} $\delta g_{\mu\nu}$ denoting the linear perturbation of the metric and $\bar{g}_{\mu\nu}$ the background metric. From now on, we do not use bars to denote the background components as we will not write the full metric explicitly and there is no risk of mixing them up.

Given $\textrm{d}s^2 = (g_{\mu\nu} + \delta g_{\mu\nu}) \textrm{d}x^{\mu} \textrm{d}x^{\nu}$, we can write the line element containing only the scalar part of the metric perturbation as \cite{Lyth:2009zz}: \begin{equation}\label{eq:pert-line-element-flrw-lapsefunc} \textrm{d}s^2 = -\mathcal{N}^2(1+2A) \textrm{d}t^2-2\partial_i B \textrm{d}x^{i} \textrm{d}t+a^2 \left[(1-2\psi)\delta_{ij}+2\partial_i \partial_j E\right]\textrm{d}x^{i} \textrm{d}x^{j}~,\end{equation} where $A(x)$ is the linear perturbation of the lapse function, $\partial_i B(x)$ the shift vector perturbation, $\psi(x)$ the intrinsic curvature perturbation and $\partial_i \partial_j E(x)$ the anisotropy in the spatial metric. Under the conformal transformation \eqref{eq:conf-metric}, we obtain the linear perturbation components of the metric in the `hatted' frame: \begin{equation}\label{eq:conf-relation-perturbation-metric} \delta \hat{g}_{\mu\nu} = \Omega^2 \left(\delta g_{\mu\nu}+2\frac{\delta \Omega}{\Omega} g_{\mu\nu}\right)~,\end{equation} where $\Omega^2$ denotes the background conformal factor. The same expression in Eq.~\eqref{eq:perturbed-metric-generic} applies to the frame with carets. Thus, using Eq.~\eqref{eq:conf-relation-perturbation-metric}, the line element in that frame is: \begin{align}\nonumber&\textrm{d}\hat{s}^2 =-\left(1+2A+2\frac{\delta \Omega}{\Omega}\right)\Omega^2\mathcal{N}^2 \textrm{d}t^2-2\Omega^2 \partial_i B \textrm{d}x^{i} \textrm{d}t+\Omega^2 a^2 \left[\left(1-2\psi +2\frac{\delta \Omega}{\Omega}\right)\delta_{ij}+\right.\\
&\left.+2\partial_i \partial_j E\right]\textrm{d}x^i \textrm{d}x^{j}~,\end{align} and we identify the following perturbation variables in the new frame: \begin{align}\label{eq:conf-trans-A}&\hat{A} \equiv A+\frac{\delta \Omega}{\Omega}~,\\
&\hat{B} \equiv \Omega^2 B~,\\
\label{eq:conf-trans-psi}&\hat{\psi} \equiv \psi -\frac{\delta \Omega}{\Omega}~,\\
\label{eq:conf-trans-E}&\hat{E} \equiv E~.\end{align} 

\subsection{Gauge transformations}
In addition to the conformal transformations we introduced in the previous section, we consider linear gauge transformations \cite{Baumann:2009ds}: \begin{equation} x^{\mu} \rightarrow x^{\mu} + \xi^{\mu}(x)~,\end{equation} where the gauge transformation vector is parametrised as: \begin{equation} \xi^{\mu}(x) = (\alpha(x), \partial^{i} \beta(x))~,\end{equation} for scalar fluctuations. Under these gauge transformation, the perturbations of the metric \eqref{eq:pert-line-element-flrw-lapsefunc} transform as: \begin{align}\label{eq:gauge-trans-A}&A \rightarrow A-\dot \alpha -\frac{\dot{\mathcal{N}}}{\mathcal{N}} \alpha~,\\
\label{eq:gauge-trans-B}&B \rightarrow B-\mathcal{N}^2 \alpha +a^2 \dot \beta~,\\
\label{eq:gauge-trans-psi}&\psi \rightarrow \psi + H \mathcal{N}\alpha~,\\
\label{eq:gauge-trans-E}& E\rightarrow E-\beta~.\end{align} These represent gauge transformation rules, see e.g. Refs.~\cite{Lyth:2009zz,Carroll:1997ar,Hwang:2001qk}, that we generalised to include a generic lapse function $\mathcal{N}$.

Besides the metric, another important tensor is the energy-momentum one. The scalar components of the linear perturbation of $\tensor{T}{^{\mu}_{\nu}}$ are \cite{Lyth:2009zz}: \begin{align}&\delta \tensor{T}{^{0}_{0}} = -\delta \rho~,\\
\label{eq:momentum-pert-0i-EM-tensor}&\delta \tensor{T}{^{0}_{i}} = \partial_i \Psi~,\\
&\delta \tensor{T}{^{i}_{j}} = \delta P \delta^{i}_{j} + \tensor{\Pi}{^{i}_{j}}~,\end{align} where $\tensor{\Pi}{^{i}_{j}}$ is the traceless part of $\delta \tensor{T}{^{i}_{j}}$ known as anisotropic pressure \cite{Baumann:2009ds}. $\rho$ and $P$ are the homogeneous energy density and isotropic pressure respectively, and $\delta \rho$ and $\delta P$ are their corresponding linear perturbations. $\Psi$ is the linear scalar momentum perturbation. It can be shown that these perturbations transform as \cite{Baumann:2009ds}: \begin{align}&\delta \rho \rightarrow \delta \rho -\dot \rho \alpha~,\\
\label{eq:gauge-trans-Psi}&\Psi \rightarrow \Psi +(\rho + P)\alpha~,\\
&\delta P \rightarrow \delta P -\dot P \alpha~,\\
&\tensor{\Pi}{^{i}_{j}} \rightarrow \tensor{\Pi}{^{i}_{j}}~,\end{align} when subject to the gauge transformations.  

\subsection{Scalar-tensor theory}\label{sec:scalar-tensor-theory-explained}
Having reviewed both the conformal and gauge transformations, we now proceed to introduce the specific model of modified gravity; that is, $F(\phi)R$ gravity.

For a canonical scalar field $\phi$ that couples non-minimally to gravity, the action in the Jordan frame is: \begin{equation}\label{eq:jordan-action} S = \int \textrm{d}^4x \sqrt{-g} \left[\frac{1}{2} F(\phi) R -\frac{1}{2} g^{\mu\nu} \partial_{\mu} \phi \partial_{\nu} \phi - V(\phi)\right]~,\end{equation} where $V(\phi)$ is the potential of the field and $F(\phi)$ the coupling function. In this frame, the gravitational sector of the action includes a scalar degree of freedom $\phi$. This action can be brought into the Einstein frame form by performing a conformal rescaling of the metric (see Eq.~\eqref{eq:conf-metric}) and choosing the conformal factor such that \cite{Maeda:1988ab} (remember that $M^2_{\mathrm{Pl}} =1$ so that $F$ is dimensionless): \begin{equation} \Omega^2 = F(\phi)~.\end{equation} The Ricci scalar transforms as \cite{Dabrowski:2008kx}: \begin{equation} R = F \left[\hat{R}+\frac{3}{F}\left(\hat{\Box} F -\frac{3}{2F}\hat{g}^{\mu\nu} \partial_{\mu} F \partial_{\nu}F\right)\right]~,\end{equation} where $\hat{\Box} \equiv \hat{g}^{\mu\nu} \hat{\nabla}_{\mu} \hat{\nabla}_{\nu}$ is the d'Alembert operator and $\hat{\nabla}_{\mu}$ the covariant derivative with Levi-Civita connection associated with the rescaled metric $\hat{g}_{\mu\nu}$. Then, the action in the Einstein frame reads: \begin{equation} \label{eq:einstein-action}S = \int \textrm{d}^4x \sqrt{-\hat{g}} \left[\frac{1}{2} \hat{R}-\frac{1}{2} \hat{g}^{\mu\nu} \partial_{\mu} \varphi \partial_{\nu} \varphi -U(\varphi)\right]~,\end{equation} where the potential $U$ is given by: \begin{equation} U(\phi) = \frac{V(\phi)}{F^2(\phi)}~,\end{equation} and $\varphi$ is the canonical scalar field of the Einstein frame related to $\phi$ by \cite{Chiba:2008ia}: \begin{equation}\label{eq:field-rede} \left(\frac{\textrm{d}\varphi}{\textrm{d}\phi}\right)^2 =\mathcal{K}\equiv \frac{1}{F} \left(1+\frac{3F_{,\phi}^2}{2F}\right)~.\end{equation} We see that the gravitational sector of the action in this frame resembles that of the Einstein-Hilbert one of GR, and the scalar field is minimally coupled to gravity.

The Jordan frame action yields the following metric field equations: \begin{equation}\label{eq:metric-field-eqs-JF-action}\tensor{G}{^{\mu}_{\nu}} = \tensor{\tilde{T}}{^{\mu}_{\nu}}~,\end{equation} where $\tensor{G}{^{\mu}_{\nu}} \equiv \tensor{R}{^{\mu}_{\nu}}-\frac{1}{2} \delta^{\mu}_{\nu} R$ is the divergence-free Einstein tensor and $\tensor{R}{^{\mu}_{\nu}}$ the Ricci tensor. The covariantly conserved (by virtue of Bianchi's identity: $\nabla_{\mu} \tensor{G}{^{\mu}_{\nu}}=0$ \cite{Carroll:1997ar}) effective energy-momentum tensor $\tensor{\tilde{T}}{^{\mu}_{\nu}}$ reads: \begin{equation}\label{eq:EM-tensor-JF} \tensor{\tilde{T}}{^{\mu}_{\nu}} = \frac{1}{F} \left[\partial^{\mu} \phi \partial_{\nu}\phi +\nabla^{\mu} \partial_{\nu}F -\delta^{\mu}_{\nu}\left(\frac{1}{2} g^{\alpha\beta} \partial_{\alpha} \phi \partial_{\beta}\phi +V + \Box F\right)\right]~,\end{equation} where $\Box \equiv g^{\mu\nu} \nabla_{\mu} \nabla_{\nu}$. From the Einstein frame action, we obtain instead: \begin{equation} \tensor{\hat{G}}{^{\mu}_{\nu}} = \tensor{\hat{T}}{^{\mu}_{\nu}}~,\end{equation} where $\tensor{\hat{T}}{^{\mu}_{\nu}}$ is given by: \begin{equation}\label{eq:EM-tensor-EF} \tensor{\hat{T}}{^{\mu}_{\nu}} = \partial^{\mu} \varphi \partial_{\nu}\varphi -\delta_{\nu}^{\mu} \left(\frac{1}{2} \hat{g}^{\alpha\beta} \partial_{\alpha}\varphi \partial_{\beta}\varphi +U\right)~,\end{equation} indices being raised with $\hat{g}_{\mu\nu}$. $\tensor{\hat{G}}{^{\mu}_{\nu}} \equiv \tensor{\hat{R}}{^{\mu}_{\nu}}-\frac{1}{2} \delta^{\mu}_{\nu} \hat{R}$ is the Einstein tensor in the frame with carets, and it is divergence-free with respect to $\hat{\nabla}_{\mu}$; i.e. $\hat{\nabla}_{\mu} \tensor{\hat{G}}{^{\mu}_{\nu}} = 0$. This means that $\tensor{\hat{T}}{^{\mu}_{\nu}}$ is covariantly conserved in that frame.

Before we define the linear comoving curvature perturbation $\mathcal{R}$ and obtain the conformal relation between $\mathcal{R}$ and $\hat{\mathcal{R}}$ in Jordan and Einstein frames respectively, we outline the expressions and approximations of Ref.~\cite{Karciauskas:2022jzd} at background level. We will review those required for an eventual examination of such a relation during slow-roll inflation. 

\subsection{Slow-roll inflation in the Jordan frame} 
\label{sec:slow-roll-inflation-jordan-frame}
We now introduce the Hubble-flow (HF) parameters and define them as the relative change of the Hubble parameter (see Eq.~\eqref{eq:Hubble-param-in-terms-of-a}) as measured by the comoving observer: \begin{equation} \label{eq:def-eps-param-tau}\epsilon_{i+1} \equiv \frac{1}{H\epsilon_{i}}\frac{\textrm{d}\epsilon_{i}}{\textrm{d}\tau}~,\end{equation} where $i=1,2,...$ and:  \begin{equation}\epsilon_1 \equiv -\frac{1}{H^2} \frac{\textrm{d}H}{\textrm{d}\tau}~.\end{equation} As we pointed out below Eq.~\eqref{eq:ref-a-only-for-that}, we choose such a spacetime slicing that coordinate time equals the proper one in the Jordan frame, i.e. $\mathcal{N} = 1$. Consequently, given the Hubble parameter defined in Eq.~\eqref{eq:Hubble-param-in-terms-of-a}, this will read: \begin{equation}\label{eq:Hubb-param-Jordan-frame-adota} H = \frac{\dot a}{a}~,\end{equation} in the Jordan frame, while the HF parameters in that frame become:\begin{equation}\label{eq:HF-Jordan-frame-parameters} \epsilon_{i+1} \equiv \frac{\dot{\epsilon_i}}{H\epsilon_i}~,\end{equation} where $i=1,2,...$ and $\epsilon_1 \equiv -\dot H/H^2$. 

Besides $H$, we have a further time-dependent background scale in $F(\phi)R$ gravity, which is $\sqrt{F}$ (in GR, $\sqrt{F} = M_{\textrm{Pl}}$, or $\sqrt{F}=1$ using the geometrical units adopted in this work). We then present a second hierarchy of parameters: \begin{equation}\label{eq:def-theta-param-jordan-framee} \theta_{i+1} \equiv \frac{\dot{\theta_i}}{H\theta_i}~,\end{equation} where $i = 1,2,...$ and $\theta_1 \equiv \dot F/(2HF)$. Moreover, the equations derived in this frame are more compact if one introduces the following parameters: \begin{equation}\label{eq:def-gammaa-param} \gamma_{i+1} \equiv \frac{\dot{\gamma_i}}{H\gamma_i}~,\end{equation} where $i = 1,2,...$ and $\gamma_1^2 \equiv \mathcal{K} \dot \phi^2/(2H^2)$, and the approximations clearer if these are used in place of the HF parameters of Eq.~\eqref{eq:HF-Jordan-frame-parameters}. A useful relation for future purposes derived in Ref.~\cite{Karciauskas:2022jzd} is the one between $\epsilon_1$ and $\gamma_1^2$: \begin{equation}\label{eq:rel-gamma12-eps1} \gamma_1^2 = (\epsilon_1+\theta_1) (1+\theta_1) - \theta_1 \theta_2~.\end{equation}

Eq.~\eqref{eq:def-eps-param-tau} applies to the HF parameters of the Einstein frame as well, but with carets:\begin{equation}\label{eq:def-hateps-Einstein}\hat{\epsilon}_{i+1} \equiv \frac{1}{\hat{H} \hat{\epsilon}_i}\frac{\textrm{d}\hat{\epsilon}_i}{\textrm{d}\hat{\tau}}~,\end{equation} where $i=1,2,...$. The Hubble parameter in the Einstein frame, $\hat{H}$, can be related to $H$ of Eq.~\eqref{eq:Hubb-param-Jordan-frame-adota} by (see Eq.~\eqref{eq:Hubble-param-in-terms-of-a} and remember that, given the choice of the spacetime slicing, $\hat{\mathcal{N}} = \sqrt{F}$ in the Einstein frame): \begin{equation}\label{eq:Hubble-param-Jordan-Einstein} \hat{H} = \frac{\dot{\hat{a}}}{\sqrt{F}\hat{a}}=\frac{H}{\sqrt{F}}\left(1+\theta_1\right)~.\end{equation} 

We can write the HF parameters of the Einstein frame that are relevant at first order in the slow-roll approximation (i.e. $\hat{\epsilon}_1$ and $\hat{\epsilon}_2$ \cite{Liddle:1994dx}) as: \begin{align}&\label{eq:hat-eps1}\hat{\epsilon}_1 = \frac{\gamma_1^2}{(1+\theta_1)^2}~,\\
\label{eq:hat-eps2}&\hat{\epsilon}_2 = \frac{2}{1+\theta_1}\left(\gamma_2-\frac{\theta_1\theta_2}{1+\theta_1}\right)~,\end{align} such that they are expressed in terms of the $\gamma$ and $\theta$ parameters of the Jordan frame. Slow-roll inflation in the Einstein frame demands $\hat{\epsilon}_1 \ll 1$ and $|\hat{\epsilon}_2| \ll 1$ \cite{Liddle:1994dx}, and following on from this, it can be shown that $|\theta_1|\ll 1$, $|\epsilon_1| \ll 1$ and $\gamma_1^2 \ll 1$ \cite{Karciauskas:2022jzd}. Then, the slow-roll equations in the Jordan frame read: \begin{align}\label{eq:SR-eq1}&3H^2 \simeq \frac{V}{F}~,\\ 
\label{eq:SR-eq2}&3H \dot \phi \simeq \frac{2VF_{,\phi}-F V_{,\phi}}{F^2 \mathcal{K}}~.\end{align} These and previous formulae will be important when deriving the conformal relation of the comoving curvature perturbation, that we introduce in the next section, and to study this relation on different horizon scales and during slow-roll inflation.

\section{The comoving curvature perturbation}\label{sec3}
In view of Eqs.~\eqref{eq:gauge-trans-Psi} and \eqref{eq:gauge-trans-psi}, we may define the following gauge-invariant quantity: \begin{equation}\label{eq:comov-curv-pert-R} \mathcal{R} \equiv \psi - \frac{H \mathcal{N}}{\rho+P}\Psi=\psi-\frac{\dot a}{a}\frac{\Psi}{\rho+P}~,\end{equation} which is the linear curvature perturbation on comoving slicings (known as `comoving curvature perturbation'); i.e. those where $\delta \tensor{T}{^{0}_{i}} =0$ \cite{Hwang:1996xh,Hwang:2001fb,Gong:2011qe} or, from Eq.~\eqref{eq:momentum-pert-0i-EM-tensor}, $\Psi = 0$ (the homogeneous value of $\Psi$ vanishes by construction).

\subsection{$\mathcal{R}$ in Jordan and Einstein frames}
In the Jordan frame, $\delta \tensor{\tilde{T}}{^{0}_{i}}$ is given by (see Eq.~\eqref{eq:EM-tensor-JF}): \begin{equation}\label{eq:pert-EM-0i-comp} \delta \tensor{\tilde{T}}{^{0}_{i}} = \frac{1}{F}\partial_i \left(-\dot \phi \delta \phi -\dot{\delta F} + H \delta F + \dot F A\right)\equiv \partial_i \tilde{\Psi}~.\end{equation} One can slice the spacetime in such a way that $\tilde{\Psi} = 0$. This is what we call the comoving slicing in this work. It is important to observe that, generically, comoving slices \emph{do not} coincide with the uniform field ones.\footnote{In the case of $f(R)$ gravity, an analogous mismatch is mentioned in Ref.~\cite{Gong:2011qe}, where the uniform-$F$ slicing does not necessarily coincide with the comoving one, $F$ being defined as $F\equiv f_{,R}$.} As was pointed out in Ref.~\cite{Gong:2011qe}, perturbation quantities defined on uniform field slices are conformally invariant. It follows then that perturbations defined on the comoving slices are not generically conformally invariant.\  

Replacing the momentum perturbation in Eq.~\eqref{eq:comov-curv-pert-R} with the above expression (and given our choice of homogeneous slicing such that in the Jordan frame $\mathcal{N} =1$), the linear comoving curvature perturbation reads: \begin{equation}\label{eq:R-comov-curv-pert-Jordan-frame} \mathcal{R} = \psi +\frac{H/F}{\tilde{\rho}+\tilde{P}}\left(\dot \phi \delta \phi +\dot{\delta F} -H \delta F -\dot F A\right)~,\end{equation} where the homogeneous energy density and isotropic pressure of the effective energy-momentum tensor are $\tilde{\rho} \equiv - \tensor{\tilde{T}}{^{0}_{0}}$ and $\tilde{P} \equiv \tensor{\tilde{T}}{^{i}_{i}}/3$ respectively. The sum of the two is: \begin{equation}\label{eq:eq42} \tilde{\rho}+\tilde{P} = \frac{1}{F}\left(\dot \phi^2 -H\dot F + \ddot F\right)~.\end{equation} 

On the hypersurfaces that comove with the effective fluid in the Jordan frame, it can be shown that the curvature perturbation is conserved on large scales whenever the perturbation of the effective pressure is adiabatic \cite{Lyth:2004gb} (a result similar to what GR prescribes). To that end, we write the momentum conservation equation ($\nabla_{\mu} \tilde{T}^{\mu i} = 0$): \begin{equation} \dot{\tilde{\Psi}}+3H \tilde{\Psi} + \delta \tilde{P} +(\tilde{\rho}+\tilde{P})A = -\frac{2}{3} a^{-2} \partial_i \partial^{i} \sigma~,\end{equation} where $\sigma$ is the contribution from the anisotropic pressure \cite{Hwang:1996xh}: \begin{equation}\label{eq:def-sigma-chi} \sigma \equiv \frac{\delta F - \dot F \chi}{F}~,\end{equation} $\chi\equiv B+a^2 \dot E$ being the shear potential of worldlines orthogonal to the space-like hypersurfaces \cite{Baumann:2009ds}. We consider the momentum constraint equation too ($\tensor{G}{^{0}_i} = \tensor{\tilde{T}}{^{0}_{i}}$) \cite{Hwang:1996xh}: \begin{equation}\label{eq:momentum-const-eq-Jordan-frame} HA + \dot \psi = -\frac{1}{2}\tilde{\Psi}~.\end{equation} Using these two equations, and taking the time derivative of $\mathcal{R}$ \eqref{eq:R-comov-curv-pert-Jordan-frame}, we obtain: \begin{equation}\label{eq:time-derivative-R-v1} \dot{\mathcal{R}} = \frac{H}{\tilde{\rho}+\tilde{P}} \left(\delta \tilde{P}_{\textrm{nad}} +\frac{\dot{\tilde{P}}}{\dot{\tilde{\rho}}}\delta \tilde{\rho}^{\Psi} + \frac{2}{3} a^{-2} \partial_i \partial^{i} \sigma\right)~.\end{equation} $\delta \tilde{P}_{\textrm{nad}}$ is defined as: \begin{equation} \delta \tilde{P}_{\textrm{nad}} \equiv \delta \tilde{P}-\frac{\dot{\tilde{P}}}{\dot{\tilde{\rho}}} \delta \tilde{\rho}~,\end{equation} i.e., as the pressure perturbation on uniform-density slicings (or non-adiabatic pressure \cite{Lyth:2009zz,Baumann:2009ds}); while $\delta \tilde{\rho}^{\Psi}$: \begin{equation} \delta \tilde{\rho}^{\Psi} \equiv \delta \tilde{\rho} - 3H \tilde{\Psi}~,\end{equation} is the energy density perturbation on comoving slicings. The energy constraint equation ($\tensor{G}{^{0}_{0}} = \tensor{\tilde{T}}{^{0}_{0}}$) allows to replace $\delta \tilde{\rho}^{\Psi}$ by gradients: \begin{equation} \delta \tilde{\rho}^{\Psi} = 2a^{-2} \partial_i \partial^{i} \psi^{\chi}~,\end{equation} $\psi^{\chi}$ being the curvature perturbation on shear-free slicings ($\chi=0$): \begin{equation} \psi^{\chi} \equiv \psi + H \chi~,\end{equation} which is gauge-invariant given that $\chi$ transforms as $\chi\rightarrow \chi - \alpha$ under gauge transformations (see Eqs.~\eqref{eq:gauge-trans-B} and \eqref{eq:gauge-trans-E} and the definition of $\chi$ below Eq.~\eqref{eq:def-sigma-chi}). Eq.~\eqref{eq:time-derivative-R-v1} then becomes: \begin{equation} \dot{\mathcal{R}} = \frac{H}{\tilde{\rho}+\tilde{P}} \left[\delta \tilde{P}_{\textrm{nad}} +2a^{-2} \partial_i \partial^{i} \left(\frac{1}{3} \sigma +\frac{\dot{\tilde{P}}}{\dot{\tilde{\rho}}}\psi^{\chi} \right) \right]~.\end{equation} 

The reason why we obtain an equation for $\dot{\mathcal{R}}$ similar to the one from GR (which can be found in Ref.~\cite{Weinberg:2003sw}, for example) is because the fluid that is chosen to define the comoving slicing is the one whose energy-momentum tensor is covariantly conserved (the effective fluid) \cite{Lyth:2004gb}. In the Einstein frame, this effective fluid corresponds to that of the canonical scalar field $\varphi$ (see Eq.~\eqref{eq:EM-tensor-EF}). From Eq.~\eqref{eq:EM-tensor-EF}, we calculate $\delta \tensor{\hat{T}}{^{0}_{i}}$, which is: \begin{equation}\label{eq:0i-component-Einstein-frame}\delta \tensor{\hat{T}}{^{0}_{i}}= -\frac{1}{F}\dot \varphi \partial_i \delta \varphi\equiv \partial_i \hat{\Psi}~.\end{equation} The comoving curvature perturbation in this case is (remember that $\hat{\mathcal{N}} = \sqrt{F})$: \begin{equation}\label{eq:R-comov-curv-pert-Einstein-frame} \hat{\mathcal{R}} = \hat{\psi} +\frac{\hat{H}}{\textrm{d}\varphi/\textrm{d}\hat{\tau}}\delta \varphi~,\end{equation} where we have used the fact that $\hat{\rho}+\hat{P}$ is: \begin{equation} \hat{\rho} + \hat{P} = -\tensor{\hat{T}}{^{0}_{0}} + \tensor{\hat{T}}{^{i}_{i}}/3=\dot \varphi^2/F~.\end{equation} 

\subsection{The conformal equivalence of $\mathcal{R}$}
\label{sec:conf-equiv-R-comov}
As can be seen from Eq.~\eqref{eq:0i-component-Einstein-frame} above, the comoving curvature perturbation coincides with the curvature perturbation on uniform-$\varphi$ slicings in the Einstein frame $\hat{\psi}^{\delta \varphi}$ (i.e., if the perturbation of the canonical scalar field is set to zero, $\delta \varphi = 0$, then the scalar momentum perturbation vanishes: $\delta \tensor{\hat{T}}{^{0}_{i}}=0$). This coincidence is due to the fact that the covariantly conserved, effective energy-momentum tensor in this frame is that of the canonical scalar field $\varphi$, as was mentioned above. Given Eq.~\eqref{eq:R-comov-curv-pert-Einstein-frame}, we easily verify that $\hat{\psi}^{\delta \varphi}$ is invariant under the conformal transformation and the field redefinition of Eq.~\eqref{eq:field-rede} \cite{Chiba:2008ia}: \begin{equation}\label{eq:conf-trans-psi-deltaphi} \hat{\mathcal{R}}=\hat{\psi}^{\delta \varphi} = \hat{\psi}+\frac{\hat{H}}{\textrm{d}\varphi/\textrm{d}\hat{\tau}}\delta \varphi = \psi - \frac{\delta F}{2F}+\frac{H}{\dot \phi}(1+\theta_1) \delta \phi = \psi + \frac{H}{\dot \phi}\delta \phi = \psi^{\delta \phi}~,\end{equation} where Eq.~\eqref{eq:conf-trans-psi} was used to replace $\hat{\psi}$ by $\psi$, and we remind the reader that the Einstein frame is obtained by setting $\Omega^2 = F$ (see Sec.~\ref{sec:scalar-tensor-theory-explained}). Also, $\delta \varphi = \sqrt{\mathcal{K}}\delta \phi$ (see Eq.~\eqref{eq:field-rede}) and $\delta F = F_{,\phi} \delta \phi$. Eq.~\eqref{eq:conf-trans-psi-deltaphi} is generic, valid at all scales and irrespective of how the background spacetime behaves. However, in the Jordan frame, an observer that comoves with the canonical scalar field $\phi$ does not observe vanishing effective momentum perturbation. The exact relation between $\mathcal{R}$ and $\psi^{\delta \phi}$ can be derived from Eqs.~\eqref{eq:R-comov-curv-pert-Jordan-frame} and \eqref{eq:momentum-const-eq-Jordan-frame}: \begin{equation}\label{eq:exact-rel-R-and-psideltaphi} \mathcal{R} = \psi^{\delta \phi} + \frac{\theta_1}{\epsilon_1(1+\theta_1)}\frac{\dot \psi^{\delta \phi}}{H}~.\end{equation} The second term is the direct consequence of the fact that uniform-field slices, where $\delta \phi = 0$, do not coincide with comoving slices, as it was discussed below Eq.~\eqref{eq:pert-EM-0i-comp}. Since the perturbed quantities defined on the uniform-field slices are conformally invariant (see Ref.~\cite{Gong:2011qe}), it is not surprising to find that $\mathcal{R} \neq \hat{\mathcal{R}}$.\ 

We then proceed to derive the corresponding relation between $\mathcal{R}$ and $\hat{\mathcal{R}}$ for a generic $F(\phi)R$ gravity theory at linear order and check if the invariance may be true under certain conditions. We begin by writing the comoving curvature perturbation in the Jordan frame as (see Eq.~\eqref{eq:R-comov-curv-pert-Jordan-frame}): \begin{equation} \mathcal{R} = \frac{H/F}{\tilde{\rho}+\tilde{P}}\left(\dot \phi \delta \phi^{\psi} +\dot{\delta F^{\psi}}-H \delta F^{\psi} -\dot F A^{\psi}\right)~,\end{equation} where we defined the following gauge-invariant quantities in that frame: \begin{align}&\delta \phi^{\psi} \equiv \delta \phi +\frac{\dot \phi}{H}\psi~,\\ 
&\delta F^{\psi} \equiv \delta F + \frac{\dot F}{H} \psi~,\\
&A^{\psi} \equiv A +\left(\frac{\psi}{H} \right)^{\bullet}~,\end{align} and used Eq.~\eqref{eq:eq42}. From the momentum constraint equation (Eq.~\eqref{eq:momentum-const-eq-Jordan-frame}), we obtain $A^{\psi}$: \begin{equation} A^{\psi} = \frac{1}{2HF}\frac{\dot \phi \delta \phi^{\psi} +\dot{\delta F^{\psi}}-H \delta F^{\psi}}{1+\theta_1}~.\end{equation} Hence: 
\begin{equation}\mathcal{R} = \frac{H}{\dot \phi} \frac{1}{1+\theta_1} \frac{1}{\epsilon_1} \left\{\left[\epsilon_1 \left(1+\theta_1 \right) +\theta_1 \left(\theta_{\mathcal{K}}-\gamma_2 \right) \right] \delta \phi^{\psi} +\theta_1 \frac{\dot{\delta \phi^{\psi}}}{H} \right\}~,\end{equation} where we used $\delta F = F_{,\phi} \delta \phi$ and introduced the parameter: \begin{align}&\theta_{\mathcal{K}}\equiv \frac{\dot{\mathcal{K}}}{2H\mathcal{K}}~.\end{align} We have used $\tilde{\rho} + \tilde{P} = -2\dot H$ too (which can be derived from the homogeneous metric field equations \eqref{eq:metric-field-eqs-JF-action}), and the first HF parameter $\epsilon_1$ from Eq.~\eqref{eq:HF-Jordan-frame-parameters}. Since $\hat{\mathcal{R}} = \hat{\psi}^{\delta \varphi} = \psi^{\delta \phi}$ (see Eq.~\eqref{eq:conf-trans-psi-deltaphi}), we have: \begin{equation}\mathcal{R} = \hat{\mathcal{R}} \left[1+\frac{\theta_1}{\epsilon_1\left(1+\theta_1 \right)}\left(\theta_{\mathcal{K}}-\gamma_2+\frac{\dot{\delta \phi^{\psi}}}{H \delta \phi^{\psi}}\right)\right]~.\end{equation} 

Now, given that: \begin{equation} \delta \phi^{\psi} =\frac{\dot \phi}{H} \psi^{\delta \phi} = \frac{\dot \phi}{H} \hat{\mathcal{R}} =-\frac{\dot \phi}{H\hat{z}}\hat{u}= \frac{1+\theta_1}{a\sqrt{F\mathcal{K}}}\hat{u}~,\end{equation} such that $\hat{u}\equiv -\hat{z}\hat{\mathcal{R}}$, where $\hat{z}$ is the Mukhanov variable in the Einstein frame \cite{Baumann:2009ds}, defined by: \begin{equation} \hat{z} \equiv -\frac{\hat{a}}{\hat{H}}\frac{\textrm{d} \varphi}{\textrm{d}\hat{\tau}}~,\end{equation} then: \begin{equation} \frac{\dot{\delta \phi^{\psi}}}{H \delta \phi^{\psi}} = -1-\theta_1-\theta_{\mathcal{K}}+\frac{\theta_1 \theta_2}{1+\theta_1} +\frac{\dot{\hat{u}}}{H\hat{u}}~.\end{equation} So we arrive at: \begin{equation}\label{eq:first-version-conf-relv1} \mathcal{R} = \hat{\mathcal{R}} \left[1-\frac{\theta_1}{\epsilon_1}\left(1+\frac{1}{2} \hat{\epsilon}_2 -\frac{1}{\hat{H} \hat{u}}\frac{\textrm{d}\hat{u}}{\textrm{d}\hat{\tau}}\right)\right]~,\end{equation} where we used Eqs.~\eqref{eq:Hubble-param-Jordan-Einstein} and \eqref{eq:hat-eps2} too.\ 

As we argued at the end of Sec.~\ref{sec:slow-roll-inflation-jordan-frame}, slow-roll inflation demands that $\hat{\epsilon}_1, |\hat{\epsilon}_2| \ll 1$, and from these it can be obtained that $|\theta_1| \ll 1$. Hence, during slow-roll, Eq.~\eqref{eq:first-version-conf-relv1} can be written as: \begin{equation}\label{eq622} \mathcal{R} \simeq \hat{\mathcal{R}} \left[1-\frac{\theta_1}{\epsilon_1} \left(1-\frac{1}{\hat{H}\hat{u}}\frac{\textrm{d}\hat{u}}{\textrm{d}\hat{\tau}}\right)\right]~,\end{equation} and $H\simeq \sqrt{F} \hat{H}$ as $|\theta_1| \ll 1$ (see Eq.~\eqref{eq:Hubble-param-Jordan-Einstein}). $\hat{u}$ obeys the Mukhanov-Sasaki equation in Fourier space during slow-roll \cite{Baumann:2009ds}: \begin{equation}\label{eq:mukhanov-sasaki-eq} \hat{u}_k^{''} +\left(k^2-\frac{2}{\eta^2} \right) \hat{u}_k = 0~.\end{equation} Primes in this equation are used to denote derivatives with respect to the conformal time, defined as: \begin{equation} \textrm{d}\eta \equiv \frac{\mathcal{N}}{a}\textrm{d}t~.\end{equation} We see that the conformal time is  invariant under conformal transformation given Eqs.~\eqref{eq:ref-N-only-for-that} and \eqref{eq:ref-a-only-for-that}. Also, the comoving wavenumber $k$ is invariant as well because the spatial coordinates are not affected by the conformal rescaling of the metric. 

The solution to Eq.~\eqref{eq:mukhanov-sasaki-eq} is \cite{Baumann:2009ds}: \begin{equation}\label{eq:sol-to-mukhanov-sasaki} \hat{u}_k(\eta) = \frac{e^{-ik\eta}}{\sqrt{2k}} \left(1-\frac{i}{k\eta}\right)~,\end{equation} once the Bunch-Davies vacuum has been imposed as initial condition \cite{Baumann:2009ds,Lyth:2005ze}. 

The conformal time $\eta$ is related to the conformal Hubble parameter in the Einstein frame, $\hat{\mathcal{H}} \equiv \hat{a}\hat{H}$, by $\eta \simeq -1/\hat{\mathcal{H}}$ if slow-roll holds\footnote{Strictly speaking, the equality holds only for exact de Sitter (constant $\hat{H}$). However, given that the usual assumption is that slow-roll is close to de Sitter (quasi-de Sitter), this relation can still be used.} \cite{Baumann:2009ds}. Hence: \begin{equation}\label{eq:approx-uk-v1}1-\frac{1}{\hat{H} \hat{u}_k}\frac{\textrm{d}\hat{u}_k}{\textrm{d}\hat{\tau}}=1-\frac{\hat{u}_k^{'}}{\hat{\mathcal{H}}\hat{u}_k}= \frac{k^2}{\hat{\mathcal{H}}^2} \frac{1}{1-ik/\hat{\mathcal{H}}}~.\end{equation} Eq.~\eqref{eq622} becomes: \begin{equation} \mathcal{R}_k \simeq \frac{1-ik/\hat{\mathcal{H}}-\omega (k/\hat{\mathcal{H}})^2}{1-ik/\hat{\mathcal{H}}}\hat{\mathcal{R}}_k ~,\end{equation} in Fourier space, where: \begin{equation}\label{eq:def-omega-ratio-scales}\omega \equiv \frac{\theta_1}{\epsilon_1}=-\frac{\textrm{d}\ln \sqrt{F}}{\textrm{d}\ln H}~,\end{equation} $\epsilon_1$ and $\theta_1$ being defined after Eqs.~\eqref{eq:HF-Jordan-frame-parameters} and \eqref{eq:def-theta-param-jordan-framee} respectively. $\omega$ measures how large the variation of $\ln H$ and $\ln \sqrt{F}$ is relative to each other over the number of e-folds $N\equiv \ln a$. If $|\omega| <1$, $\ln H$ varies faster than $\ln \sqrt{F}$, while $|\omega|>1$ means that the rate of change of $\ln H$ is slower than that of $\ln \sqrt{F}$. We emphasise again that $H$ and $\sqrt{F}$ are the two time-dependent background scales of the theory in the Jordan frame. Notice that if $\omega = 0$ (because $\theta_1 = 0$, as in GR), then $\mathcal{R}_k = \hat{\mathcal{R}}_k$ for all $k$, as expected. 

On the superhorizon regime ($k/\hat{\mathcal{H}}\ll 1$), we have: \begin{equation}\label{eq:superhorizon-scales-comov-curv-pert-conf-rel} \mathcal{R}_{k\ll \hat{\mathcal{H}}} \simeq \left[1-\omega\left(\frac{k}{\hat{\mathcal{H}}}\right)^2\right]\hat{\mathcal{R}}_{k\ll \hat{\mathcal{H}}}~,\end{equation} and the equivalence of the two linear comoving curvature perturbations in Jordan and Einstein frames is valid up to a term quadratic in gradients. We notice that this equivalence implies that the comoving and uniform-$\phi$ slicings coincide whenever $|\omega|k^2/\mathcal{\hat{H}}^2\ll 1$ in slow-roll (see Eq.~\eqref{eq:conf-trans-psi-deltaphi}): \begin{equation} \psi^{\delta \phi}_{|\omega|k^2\ll \hat{\mathcal{H}}^2} = \hat{\psi}^{\delta \varphi}_{|\omega|k^2\ll \hat{\mathcal{H}}^2} = \hat{\mathcal{R}}_{|\omega|k^2\ll \hat{\mathcal{H}}^2} \simeq \mathcal{R}_{|\omega|k^2\ll \hat{\mathcal{H}}^2}~.\end{equation}   

For subhorizon modes instead ($k/\hat{\mathcal{H}}\gg 1$): \begin{equation}\label{eq:subhorizon-scales-comov-curv-pert-conf-rel}\mathcal{R}_{k\gg \hat{\mathcal{H}}} \simeq\left(1-i \omega \frac{k}{\hat{\mathcal{H}}}\right) \hat{\mathcal{R}}_{k\gg \hat{\mathcal{H}}}~,\end{equation} and the equivalence between $\mathcal{R}$ and $\hat{\mathcal{R}}$ in the two frames does not hold. However, due to the additional factor $\omega$, the length scales on which $\mathcal{R}$ can be considered to be conformally invariant depends on the time dependence of $F$ and $H$. For models with very small $\omega$ such an invariance already holds on subhorizon scales. For very large $\omega$, it holds only long after such scales exit the horizon.\ 

When discussing the superhorizon and subhorizon regimes, we delineated them by the magnitude of the $k/(aH)$ ratio. However, the equalities $k=aH$ and $k = \hat{a} \hat{H}$ are satisfied on two different homogeneous slices, as can be seen from Eq.~\eqref{eq:Hubble-param-Jordan-Einstein} (see also Ref.~\cite{Karciauskas:2022jzd}): \begin{equation}\label{eq:aHhathatless-coincidence} \hat{a} \hat{H} = aH (1+\theta_1)~.\end{equation} Fortunately, since the analysis performed in this section relies on the slow-roll approximation, where $|\theta_1| \ll 1$, the difference between those two slices is immaterial. Notice that this implies the equivalence between the conformal Hubble parameters in the two frames, $\mathcal{H} \simeq \hat{\mathcal{H}}$, during slow-roll, where $\mathcal{H}\equiv a H$ in the Jordan frame ($\hat{\mathcal{H}}$ was defined after Eq.~\eqref{eq:sol-to-mukhanov-sasaki}) 

\subsection{Conformal invariance in generalised induced gravity models}
As we mentioned in Sec.~\ref{sec:slow-roll-inflation-jordan-frame}, the $\gamma$ parameters make approximations clearer if they replace the HF parameters of the Jordan frame. Assuming the slow-roll approximation, Eq.~\eqref{eq:rel-gamma12-eps1} may be written as: \begin{equation}\label{eq:approx-rel-slowrollgamma} \gamma_1^2 \simeq \epsilon_1+\theta_1~,\end{equation} where we have considered $|\theta_2| \ll 1$ for simplicity, although it was pointed out in Ref.~\cite{Karciauskas:2022jzd} that the slow-roll assumption still permits $|\theta_2| \sim 1$. Given the slow-roll relation between $\gamma_1^2$ and $\epsilon_1$, $\omega$ can be written as:\begin{equation} \omega \simeq \frac{\theta_1}{\gamma_1^2-\theta_1} = \left(\frac{\gamma_1^2}{\theta_1}-1\right)^{-1}=\left(\frac{\mathcal{K}\dot \phi}{H}\frac{F}{F_{,\phi}}-1\right)^{-1}~.\end{equation} 

Now, two different regimes are in order: either $\mathcal{K} \simeq 1/F$ because $F_{,\phi}^2 \ll F$; or $\mathcal{K} \simeq 3F_{,\phi}^2/(2F^2)$ as $F_{,\phi}^2 \gg F$. We explored both possibilities in Ref.~\cite{Karciauskas:2022jzd}. The latter yields: \begin{equation}\label{eq:omega-minus-1} \omega \simeq \left(3\theta_1-1\right)^{-1} \simeq -1~,\end{equation} so $\omega \sim \mathcal{O}(1)$ (i.e. the evolution rate of both scales, $H$ and $\sqrt{F}$, is similar; see Eq.~\eqref{eq:def-omega-ratio-scales}). On the other hand, $F_{,\phi}^2 \ll F$ yields: \begin{equation} \omega \simeq \left(\frac{\dot \phi}{HF_{,\phi}}-1\right)^{-1}\simeq \left(1-\frac{FV_{,\phi}}{F_{,\phi}V}\right)^{-1}~,\end{equation} where the slow-roll equations, Eqs.~\eqref{eq:SR-eq1} and \eqref{eq:SR-eq2}, were used. $\omega$ would be larger than unity if: \begin{equation}\label{eq:app-alpha-largev1} \frac{F_{,\phi}}{F} \simeq \frac{V_{,\phi}}{V}~.\end{equation} This is viable given that: \begin{equation} \frac{1}{2} \left(\frac{V_{,\phi}}{V}\right)^2 \ll \frac{1}{F}~,\end{equation} must be satisfied if $\hat{\epsilon}_1 \ll 1$ (see Ref.~\cite{Karciauskas:2022jzd} for more details), and if we insert Eq.~\eqref{eq:app-alpha-largev1} we obtain $F_{,\phi}^2 \ll F$, which is precisely the condition we had imposed in this case. For generalised models of induced gravity inflation\footnote{See Refs.~\cite{Accetta:1985du,Fakir:1992cg,Kaiser:1993bq,Kaiser:1994wj} for more details about the initial development of the induced gravity inflationary model.} however, it was shown in Ref.~\cite{Karciauskas:2022jzd} that those cases, in which $F_{,\phi}^2 \ll F$, do not fall within the $2\sigma$ region of the latest BICEP/Keck results \cite{BICEP:2021xfz}, in contrast to those satisfying $F_{,\phi}^2 \gg F$. Therefore, for this class of models, $\omega \sim \mathcal{O}(1)$, and the comoving curvature perturbation is invariant under the conformal transformation for a few e-folds after horizon-crossing (see Eq.~\eqref{eq:superhorizon-scales-comov-curv-pert-conf-rel}).

\subsection{Inflation observables}
\label{sec:inflation-obs-section}
In Sec.~\ref{sec:conf-equiv-R-comov}, we have shown the conformal invariance of the linear perturbation $\mathcal{R}$ on sufficiently large scales in slow-roll using the conformal transformation. Now we are going to check the consistency of that conformal invariance by doing all the calculations in the Jordan frame. We will calculate the inflation observables in that frame and compare them with those obtained in Ref.~\cite{Karciauskas:2022jzd}, where we considered the Einstein frame expressions in the first place and then applied the conformal transformation.  

We begin with the exact relation between $\mathcal{R}$ and $\psi^{\delta \phi}$ at linear order \eqref{eq:exact-rel-R-and-psideltaphi}. $\psi^{\delta \phi}$ can be written as: \begin{equation}\psi^{\delta \phi} = \frac{H}{\dot \phi}\delta \phi^{\psi} \equiv -\frac{u}{z}~,\end{equation} where we defined $u$ and $z$ as: \begin{align} &u \equiv \frac{\sqrt{F\mathcal{K}}}{1+\theta_1}a \delta \phi^{\psi}~,\\
&z \equiv -a\frac{\sqrt{2F\gamma_1^2}}{1+\theta_1}~,\end{align} respectively. Hence, Eq.~\eqref{eq:exact-rel-R-and-psideltaphi} may be rewitten as: \begin{equation}\mathcal{R} = -\frac{u}{z} \left[1-\frac{\omega}{1+\theta_1}\left(1-\frac{u^{'}}{\mathcal{H}u}+\theta_1+\gamma_2-\frac{\theta_1\theta_2}{1+\theta_1}\right)\right]~,\end{equation} where $\omega$ was defined in Eq.~\eqref{eq:def-omega-ratio-scales} and $z^{'}/z$ is: \begin{equation}\frac{z^{'}}{z} = \mathcal{H} \left(1+\theta_1+\gamma_2-\frac{\theta_1 \theta_2}{1+\theta_1}\right)~.  \end{equation}    

$u$ can be shown to satisfy an equation similar to Mukhanov-Sasaki equation \cite{Hwang:1996xh}: \begin{equation}\label{eq:analogue-of-MS-equation-FPHIR} u_k^{''} + \left(k^2 -\frac{z^{''}}{z}\right)u_k = 0~,\end{equation} in Fourier space, where: \begin{align}\nonumber&\frac{z^{''}}{z} = \mathcal{H}^2 \left[(1+\theta_1+\gamma_2)\left(2+\theta_1+\gamma_2-\epsilon_1-2\frac{\theta_1\theta_2}{1+\theta_1}\right)+\gamma_2\gamma_3+\frac{\theta_1\theta_2}{1+\theta_1} \left(\theta_1+\epsilon_1-\right.\right.\\
&\left.\left.-\theta_2-\theta_3+2\frac{\theta_1\theta_2}{1+\theta_1}\right)\right]~.\end{align} During slow-roll, assuming that all the parameters are negligible, we have $z^{''}/z \simeq 2/\eta^2$ and we can use the same solution for $u_k$ as that in Eq.~\eqref{eq:sol-to-mukhanov-sasaki}. Then (see Eq.~\eqref{eq:approx-uk-v1}): \begin{equation}1-\frac{u_k^{'}}{\mathcal{H}u_k} = \frac{k^2}{\mathcal{H}^2} \frac{1}{1-ik/\mathcal{H}}~. \end{equation} Hence, whenever $|\omega|k^2/\mathcal{H}^2\ll 1$, $\mathcal{R}_{|\omega| k^2\ll \mathcal{H}^2} \simeq -u_{|\omega| k^2\ll \mathcal{H}^2}/z= \psi^{\delta \phi}_{|\omega|k^2\ll \mathcal{H}^2}$ during slow-roll. 

Given that result, we may write the power spectrum of the linear comoving curvature perturbation, $A_s$, as \cite{Baumann:2009ds} (we drop `$|\omega|k^2\ll \mathcal{H}^2$' for simplicity): \begin{equation}\label{eq:amplitude-of-scalar-pert-v1}A_s \equiv \frac{k^3}{2\pi^2} \left|\mathcal{R}_k\right|^2 \simeq \frac{k^3}{2\pi^2}\frac{\left|u_k\right|^2}{z^2} \simeq \frac{H^2}{8\pi^2F\gamma_1^2} \simeq \frac{V}{24\pi^2F^2 \gamma_1^2}~.\end{equation} In the last equality, one of the two slow-roll equations, Eq.~\eqref{eq:SR-eq1}, shown in Sec.~\ref{sec:slow-roll-inflation-jordan-frame}, was used. 

Another important inflation observable is the scalar spectral index $n_s$, which can be written as \cite{Baumann:2009ds}: \begin{equation} n_s -1 \equiv \frac{\textrm{d} \ln A_s}{\textrm{d}\ln k}~.\end{equation} To calculate the spectral index, we use the chain rule first \cite{Baumann:2009ds}: \begin{equation}\frac{\textrm{d}}{\textrm{d}\ln k}= \frac{\textrm{d}N}{\textrm{d}\ln k}\frac{\textrm{d}}{\textrm{d}N}= \frac{1}{1-\epsilon_1} \frac{\textrm{d}}{\textrm{d}N}~,\end{equation} where we have used the horizon-crossing relation, $k=aH$, $N\equiv \ln a$ or $\textrm{d} N = H \textrm{d}t$ being the number of e-folds as defined below Eq.~\eqref{eq:def-omega-ratio-scales}. Hence (given that $|\epsilon_1| \ll 1$ during slow-roll): \begin{equation} n_s-1 \simeq \frac{\textrm{d}\ln A_s}{\textrm{d}N}\simeq 2\frac{\textrm{d}\ln H}{\textrm{d}N} -\frac{\textrm{d}\ln F}{\textrm{d}N} -\frac{\textrm{d} \ln \gamma_1^2}{\textrm{d}N} \simeq -2(\epsilon_1+\theta_1+\gamma_2)~,\end{equation} where we inserted $A_s$ from Eq.~\eqref{eq:amplitude-of-scalar-pert-v1} and in the last equality we used the definition of $\gamma_2$ that can be inferred from Eq.~\eqref{eq:def-gammaa-param}. We may replace $\epsilon_1+\theta_1$ by $\gamma_1^2$ given Eq.~\eqref{eq:approx-rel-slowrollgamma}. Hence: \begin{equation}\label{eq:ns-spectral-index-v1} n_s-1 \simeq -2(\gamma_1^2 + \gamma_2)~.\end{equation}  

Finally, the tensor-to-scalar ratio $r$ is defined as: \begin{equation}\label{eq:def-tensor-to-scalar-ratio-generic} r \equiv \frac{A_t}{A_s}~,\end{equation} where $A_t$ is the power spectrum of linear tensor perturbations with polarisation $p$: \begin{equation} A_t \equiv \frac{k^3}{2\pi^2}\sum_p \left|h_k^p\right|^2~.\end{equation} Defining: \begin{equation} v_k^{p} \equiv \frac{a}{2} \sqrt{F} h^{p}_{k}~,\end{equation} we have an equation for tensor perturbations in $F(\phi)R$ gravity similar to the Mukhanov-Sasaki equation \cite{Baumann:2009ds,Hwang:1996xh}: \begin{equation}\label{eq:eq-tensor-modes-ms} v_k^{p''} + \left(k^2 -\frac{z^{''}}{z} \right) v_k^{p} = 0~,\end{equation} where $z$ is now given by: \begin{equation} z\equiv \frac{a}{2} \sqrt{F}~.\end{equation} Hence: \begin{equation} \frac{z^{''}}{z} = \mathcal{H}^2\left(1+\theta_1\right)\left(2+\theta_1-\epsilon_1+\frac{\theta_1\theta_2}{1+\theta_1} \right)~.\end{equation} 

During slow-roll, we have $z^{''}/z\simeq 2/\eta^2$ again, and the solution to Eq.~\eqref{eq:eq-tensor-modes-ms} will resemble that of $u_k$. Therefore, on superhorizon scales: \begin{equation}\label{eq:A-t-slow-roll} A_t = \frac{k^3}{2\pi^2} \sum_{p=+,\times} \left|h_k^{p}\right|^2 = \frac{k^3}{2\pi^2} \sum_{p=+,\times} \frac{\left|v_k^{p}\right|^2}{z^2} \simeq \frac{2}{\pi^2} \frac{H^2}{F}~,\end{equation} where we summed over the two polarisation states `$+$' and `$\times$'.\

Plugging Eqs.~\eqref{eq:amplitude-of-scalar-pert-v1} and \eqref{eq:A-t-slow-roll} into Eq.~\eqref{eq:def-tensor-to-scalar-ratio-generic}, we obtain: \begin{equation}\label{eq:tensor-to-scalar-ratiov1}r \simeq 16\gamma_1^2~.\end{equation} Notice that: \begin{equation}\label{eq:equiv-at-jordan-einstein} A_t \simeq \frac{2}{\pi^2}\frac{H^2}{F} \simeq \frac{2}{\pi^2} \hat{H}^2~,\end{equation} where we used Eq.~\eqref{eq:Hubble-param-Jordan-Einstein} and the fact that, during slow-roll, $|\theta_1| \ll 1$. This is nothing but the expected amplitude of linear tensor perturbations in GR during slow-roll inflation (see e.g. Ref.~\cite{Baumann:2009ds} where $A_t$ is calculated in GR or, equivalently, the Einstein frame, and remember that $M^2_{\textrm{Pl}} = 1$). Moreover, it can be shown that the tensor perturbations are conformally invariant in general to fully non-linear order \cite{Gong:2011qe}.\    

The expressions in Eqs.~\eqref{eq:amplitude-of-scalar-pert-v1}, \eqref{eq:ns-spectral-index-v1} and \eqref{eq:tensor-to-scalar-ratiov1} are the same in Eqs.~(96)-(98) in Ref.~\cite{Karciauskas:2022jzd} respectively (except for the $\theta_1\theta_2$ term in Eq.~(97) because we took $|\theta_2| \ll 1$ here). In that reference, those equations were obtained by using the Einstein frame expression of $A_s$, $n_s$ and $r$, applying the conformal transformation and using the slow-roll approximation. Here, on the other hand, we have arrived at the same results by starting with the analogue of the Mukhanov-Sasaki equation in the Jordan frame and then using the slow-roll approximation, without resorting to the conformal transformation. This confirms the conformal invariance of the linear comoving curvature perturbation and hence the equivalence between the comoving and the uniform-field slicings on scales sufficiently large such that $|\omega| k^2/\mathcal{H}^2 \ll 1$ in slow-roll (or $|\omega| k^2/\hat{\mathcal{H}}^2\ll 1$ given that $\hat{\mathcal{H}} \simeq \mathcal{H}$ during slow-roll as explained below Eq.~\eqref{eq:aHhathatless-coincidence}).

\section{Summary and conclusions}\label{sec4}
The curvature perturbation is a central object in cosmological perturbation theory and plays an essential role in the inflationary cosmology. Although the conformal invariance of the curvature perturbation has been proved, at all scales and irrespective of the expansion rate, on the uniform-field slicing in $F(\phi)R$ gravity or, equivalently, on hypersurfaces comoving with the scalar field fluid, the same is not exactly exhibited on the slicing that comoves with the effective fluid of that theory, whose energy-momentum tensor is covariantly conserved. 

We first make it clear that the comoving and uniform-field slicings do not coincide in general. We define the comoving curvature perturbation at linear order in perturbation theory in any frame, assuming an observer that comoves with the effective fluid, which includes the gravity modifications in the Jordan frame, and derive the conformal relation between this curvature perturbation in Jordan and Einstein frames. This relation is determined on superhorizon scales up to a term which is quadratic in gradients during slow-roll (see Eq.~\eqref{eq:superhorizon-scales-comov-curv-pert-conf-rel}). It is noticed that the equation includes a term proportional to $\theta_1/\epsilon_1$. This ratio reflects the relative variation of the two time-dependent background scales of the theory: $H$ and $\sqrt{F}$ (see Eq.~\eqref{eq:def-omega-ratio-scales}), and it is found to be of order $1$ for the generalised induced gravity inflation models that match observations, discussed in Ref.~\cite{Karciauskas:2022jzd}. This means that for this class of models at least, the conformal equivalence between the Jordan and Einstein frames is ensured shortly after horizon-crossing. This result agrees with the one showed in Ref.~\cite{Gong:2011qe} non-perturbatively given that the uniform-$\phi$ and comoving slicings coincide on superhorizon scales during slow-roll inflation, as we clarified in this work. Such an equivalence cannot be asserted on the subhorizon regime however (see Eq.~\eqref{eq:subhorizon-scales-comov-curv-pert-conf-rel}).

Lastly, the same expressions for the inflation observables of Ref.~\cite{Karciauskas:2022jzd} are obtained in Sec.~\ref{sec:inflation-obs-section} using the linear comoving curvature perturbation of the Jordan frame (see Eqs.~(96)-(98) in that paper), and an equation analogous to Mukhanov-Sasaki's in GR (see Eq.~\eqref{eq:analogue-of-MS-equation-FPHIR} and Ref.~\cite{Hwang:1996xh}). To arrive at those expressions in this work, we assume the superhorizon regime and the slow-roll approximation only. No conformal rescaling of the metric is carried out, in contrast to the strategy followed in Ref.~\cite{Karciauskas:2022jzd}, where we started with the Einstein frame expressions and performed the conformal transformation to write them in terms of the Jordan frame variables. We then find it is possible to formulate those inflation observables using the comoving curvature perturbation at first order in perturbation theory in Jordan and Einstein frames. This serves as a demonstration of the conformal invariance of the linear comoving curvature perturbation on \emph{superhorizon scales} during slow-roll inflation. 

\bmhead{Acknowledgments}
M.K. is supported by the Mar\'ia Zambrano grant, provided by the Ministry of Universities from the Next Generation funds of the European Union. This work is also partially supported by the MICINN (Spain) projects PID2019-107394GB-I00/AEI/10.13039/501100011033 (AEI/FEDER, UE).

\bibliography{sn-article.bbl}

\end{document}